# Eavesdropping time and frequency: phase noise cancellation along a time-varying path, such as an optical fiber

G. Grosche

*Physikalisch-Technische Bundesanstalt, Bundesallee 100,*
*D-38116 Braunschweig, Germany*
*\*Corresponding author: gesine.grosche@ptb.de*

September 3, 2013

Single-mode optical fiber is a highly efficient connecting medium, used not only for optical telecommunications but also for the dissemination of ultra-stable frequencies or timing signals. In 1994, Ma, Jungner, Ye and Hall described a measurement and control system to deliver the same optical frequency at *two* places, namely the two ends of a fiber, by eliminating the "fiber-induced phase-noise modulation, which corrupts high-precision frequency-based applications". We present a simple detection and control scheme to deliver the same optical frequency at *many* places anywhere along a transmission path, or in its vicinity, with a relative instability of 1 part in $10^{19}$. The same idea applies to radio frequency and timing signals. This considerably simplifies future efforts to make precise timing/frequency signals available to many users, as required in some large scale science experiments.

Optical frequency references and clocks have achieved an unprecedented accuracy of better than 1 part in $10^{17}$ [1,2], with an instability near 1 part in $10^{18}$ [3]. They are formidable tools for precision experiments, since many precision measurements rely on converting the quantity to be measured into a frequency. Some of the most fundamental questions in physics relate to the quantities energy, space and time, and these quantities are directly related to frequency (and/or phase). This makes experiments probing fundamental questions accessible to frequency or phase measurements, e.g. testing for time-variations of fundamental constants [4] or large area Sagnac interferometers [5]. A prominent example is relativistic geodesy, i.e. the measurement of gravitational red-shift with optical clocks [6]. We therefore wish to transfer timing or frequency signals to other experimental sites, enabling applications outside metrology [4-7].

To date, efforts have focused on long distance connections [8-10] between just two points, one "remote" lab and one "local" lab connected by an optical fiber, using methods similar to that proposed in 1994 by Ma *et al*. [11] to correct phase perturbations between the local and remote end. For example, we have transmitted optical frequencies with a relative accuracy of $10^{-19}$ over 146 km deployed fiber [12], and remotely characterized optical clock lasers on-line with Hz-level resolution [13]. Significant efforts are now underway to establish national and even international metrology fiber networks.

One important question [14-18] is how to distribute reference frequencies to many users simultaneously in a cost-effective way. Surprisingly, with one point-to-point connection (such as a long stabilized fiber), we can "tap" this fiber anywhere and locally derive a reference frequency with the same precision as that achieved at the end-point [18]. We present the patented concept, an experimental set-up achieving relative frequency instability of $10^{-19}$, and several extensions of the idea; these include a branching design, and the multipoint dissemination of time using two-way transfer. The methods appear suitable for "small", km-scale networks, e.g. within particle accelerators or a university or city, as well as large area science projects, such as radio telescope arrays [7] or national metrology networks.

To explain the basic idea, we first consider the existing scheme for point-to-point stabilization. Fig. 1 shows a commonly implemented and well-characterized method for phase-stable transmission of an ultra-stable optical frequency $\nu_{\text{local}}$ from a local point A to a remote point Z [8-10,12,19,20]. Analogous, earlier designs enable the phase-stable transmission of radio frequency [21-22] or pulsed [23] signals. The method is reminiscent of [11,24] (though different), and can be understood by viewing the entire transmission path from A to the mirror at Z as the long arm of an interferometer.

We denote by $\Delta\phi_{ij}$ the phase shift experienced by a signal travelling from any point i to point j, and assume symmetry of the transmission path: $\Delta\phi_{ij} = \Delta\phi_{ji}$. The acousto-optic modulator AOM2 provides a fixed frequency shift to distinguish light that has reached the mirror at Z from light reflected anywhere else along the transmission path. At photodetector DetA, the returned light is superimposed with local light traveling through a short reference arm. This yields a beat signal with frequency

$$f_{\text{DetA}} = 2(f_{\text{AOM1}} + \Delta\dot\phi_{\text{AZ}}/2\pi + f_{\text{AOM2}})$$

and phase reflecting the momentary phase difference between the two interferometer arms. The beat signal $f_{\text{DetA}}$ is compared to a synthesized signal at frequency $f_{\text{synth}}$ with a phase-frequency comparator [21], or a simple mixer. A servo acts on the phase and frequency of AOM1 to maintain a constant phase difference between the beat signal and the synthesizer signal, resulting in a fixed phase relationship (modulo the frequency offset $f_{synth}/2$) between light at point Z and at point A. Thus the frequency delivered to point Z is given by $\nu_Z = \nu_{local} + f_{synth}/2 = \nu_A + f_{synth}/2$.

To generate a phase-stable signal at an intermediate point C along the transmission path, we now tap the transmitted signal in both directions. Figs. 1,2 and the analysis describe in detail one implementation [18, p.2]; similar set-ups are suitable for periodically modulated or pulsed signals [18]. Here the terminology of optical frequency transfer is used.

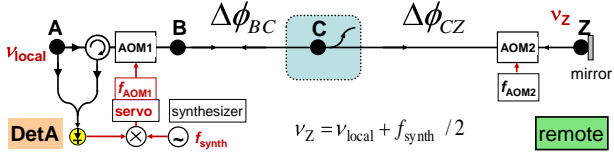

Fig. 1. Stabilized fiber link connecting the remote point Z with the local point A. For a detailed discussion, see e.g. [8,19].

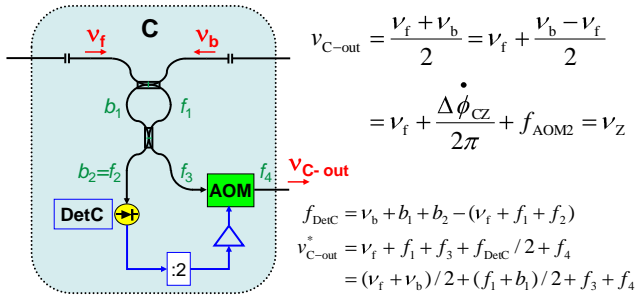

Fig. 2. Signal generation at point C, from signals tapped in forward and backward directions, with frequencies $\nu_f$ and $\nu_b$.

The forward propagating light has frequency:
$$\nu_f := \nu_{C-f} = \nu_A + f_{AOM1} + \Delta\dot\phi_{AC}/2\pi$$
whereas backward propagating light has frequency:
$$\nu_b := \nu_{C-b} = \nu_A + f_{AOM1} + \Delta\dot\phi_{AZ}/2\pi + 2f_{AOM2} + \Delta\dot\phi_{ZC}/2\pi$$
We superimpose forward and backward propagating light on photodetector DetC, generating a beat signal at frequency:

$$\begin{aligned}\nu_b - \nu_f &= \Delta\dot\phi_{CZ}/2\pi + 2f_{AOM2} + \Delta\dot\phi_{ZC}/2\pi \\ &= 2(\Delta\dot\phi_{CZ}/2\pi + f_{AOM2})\end{aligned} \quad (1)$$

The beat signal is amplified and its frequency digitally divided by two. Applying this as a correction frequency $f_{corr} := (\nu_b - \nu_f)/2$ to the forward propagating light at point C, e.g. using another AOM, we obtain a stable signal at point C: $\nu_{C-out} = \nu_f + f_{corr} = \nu_Z = \nu_{local} + f_{synth}/2$.

This can be viewed as detecting at point C the additional phase shift between points C and Z, and applying this to the signal coupled out at point C, so that its frequency and phase follow those of the signal at point Z. The output at point C is thus ideally as stable as $\nu_Z$. Since $\nu_Z = (\nu_b + \nu_f)/2$, applying $-f_{corr}$ to the *backward* propagating light at point C also yields $\nu_Z$ [18, 15].

The design has several useful properties. Many access points D, E,... etc may be operated along a single stabilized link at small extra cost for each. While [16] introduces extra frequencies, here the main link is unchanged. The signal processing is a simple and robust feed-forward system, which automatically works continuously and allows a free choice of correction bandwidth; in contrast to [16,17] it requires no additional stabilization.

Furthermore, high-power ultra-stable light can be made available [18]. If we wish to preserve optical power in the main link, asymmetric beam splitters or tap couplers extract only a few percent of the light. The extracted power may be boosted, e.g. with an erbium-doped fibre amplifier just before detector DetC. Alternatively, the weak (~µW) extracted signals are first superimposed with light from a laser at frequency $\nu_{L0}$ to give two strong heterodyne beat signals $\nu_b$-$\nu_{L0}$ and $\nu_f$-$\nu_{L0}$. Their *difference* frequency is again $\nu_b$-$\nu_f$ and is independent of $\nu_{L0}$ and its fluctuations. As before, after division by two, it may serve as a correction frequency. Alternatively, if the *mean* frequency of the two heterodyne beat signals, $(\nu_b$-$\nu_{L0} + \nu_f$-$\nu_{L0})/2$, is added as a correction to $\nu_{L0}$ (e.g. via an AOM), we obtain ultra-stable, high power laser light at frequency $\nu_L = \nu_Z$.

Key advantages of the scheme are its simple installation and its monitoring capability. By locating point Z in a second metrology lab, or even next to the input, thus forming a loop A-Z, the main link stabilization may be tested and optimized to reach the physical limits [19] and to verify its accuracy. Later, the same set-up monitors the link performance and allows online verification and assessment of the frequency distribution.

The new scheme was tested using a narrow-linewidth optical source (Koheras Adjustik fiber laser) at a frequency near 194.3 THz, on a short but intrinsically noisy fiber link. This allows exploring the fiber noise suppression and detecting system noise contributions. Our test link is a combination of a 10 m spool wound around a thin metal drum, and 100 m exposed fiber going to another laboratory (some 40 m away) and back. Point C is located at the input of the 100 m fiber. Touching or wriggling this fiber, or the fiber section wound around the drum, introduces massive phase noise, visible as a beat linewidth of several kHz. In separate experiments, initially verifying the new scheme, we also used an AOM to introduce large and well-defined perturbations.

The overall configuration is as shown in Figs. 1 and 2, but with the remote point Z and local point A located in the same laboratory. We stabilize A->Z with the standard scheme and measure the remote frequency $\nu_Z$. We also record the output frequency $\nu_{C\text{-out}}$ at the intermediate point C using the new scheme, and, additionally, the frequency correction $f_{corr}$ applied to the AOM at point C. $f_{corr}$ shows the fluctuations of the free-running link between points C and Z. All beat frequencies are recorded with totalizing counters (Kramer+Klische FXE; Π–type operation [25]), to give a time-sequence of frequency values from which we calculate the frequency instability (Allan deviation, ADEV, [25]).

Open diamonds in Fig. 3 show the frequency instability at point C when implementing the new scheme as shown in Figs. 1, 2: $\nu_{C\text{-out}}$ reaches a relative instability (ADEV in >10 kHz bandwidth) of $10^{-17}$ after 1000 s. The free-running fiber noise (full green triangles, $f_{corr}$) is clearly suppressed.

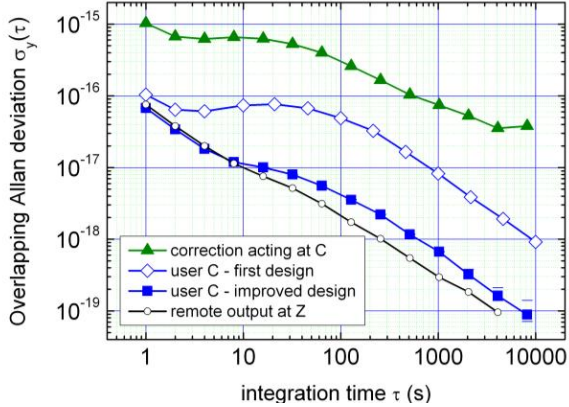

Fig. 3. Measured frequency instability (Allan deviation). Green triangles: $f_{corr}$, represents free-running fiber (see text); black open circles: $\nu_Z$, stabilized remote output at Z; blue open diamonds: $\nu_{C\text{-out}}$, stabilized output at point C, first design; full blue squares: $\nu_{C\text{-out}}$ for the improved design.

However, ADEV($\nu_{C\text{-out}}$) shows a plateau from 10…100 s, characteristic of uncompensated fiber paths [26], and the instability of $\nu_{C\text{-out}}$ is roughly 10 times that of $\nu_Z$ (small black circles). A similar, but higher, plateau of excess instability was reported in [14], using a modified implementation of this scheme [18] but operating at radio frequencies; a plateau was also seen recently in [15], reaching $4\times10^{-18}$ at 10000 s with a modified optical implementation. Our excess instability falls below $10^{-18}$ at 10000 s. Since end-point optical fiber links are feasible at the $10^{-19}$ level [9,12] even for long distances, and the newest optical clocks [1-3] achieve an instability near $10^{-18}$, we wish to eliminate this excess noise.

Analysis of the fiber paths that contribute to frequency fluctuations at point C, see Fig. 2, shows that $\nu_{C\text{-out}} = (\nu_f+\nu_b)/2 + (f_1+b_1)/2 + f_3+f_4$. [Notation: we write $f_1$ etc, for both the fiber paths and the frequency fluctuations arising from them.] The total length of exposed, uncompensated fiber is ~2-3 m.

An improved design, which minimizes uncompensated fiber paths, is shown in Fig. 4.

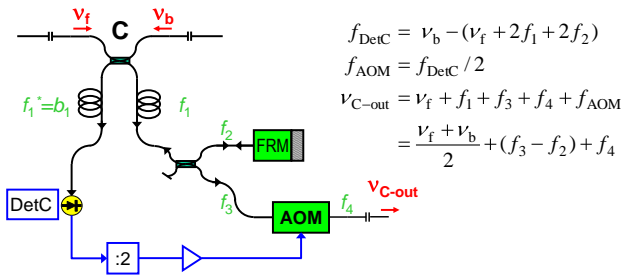

Fig. 4. Improved "tentacle" design for detection unit. FRM: Faraday rotator mirror; AOM: acousto-optic modulator; DetC: photodector. Fiber paths $f_1$ and $b_1$ contribute no noise to $\nu_{C\text{-out}}$.

For this design (full blue squares in Fig. 3), $\nu_{C\text{-out}}$ is as stable as $\nu_Z$ for short times ($10^{-17}$ at 10 s); the excess frequency instability at 1000 s is below $10^{-18}$. After 10000 s, an instability below $10^{-19}$ is reached; this is a factor 40 lower than reported in [15]. The total uncompensated fiber length (Fig. 4) was below 0.8 m, with $f_2$, $f_3$ and part of $f_4$ being co-located inside a box to minimize air currents, and mounted on a 12 mm thick aluminum board as a thermal mass. Further noise reduction is possible by environmental shielding, length-matching fibers so temperature changes enter common mode ($f_2 \sim f_3 + f_4$), and/or active temperature stabilization.

Reaching the $10^{-19}$ instability level, this proof-of-principle experiment already demonstrates that the scheme is applicable even for state-of-the-art clock comparisons. We refer to the improved version as the "tentacle" design, because the extra paths (or "tentacles") $f_1$ and $b_1$ at point C that deliver $\nu_{C\text{-out}}$ no longer introduce *any* additional noise, but cancel completely. Thus, $f_1$ and $b_1$ may be made long, enabling a true branching distribution from a fiber back-bone link, which reaches into the vicinity of the main transmission path. Further aspects of the design, relating to polarization, effects of link asymmetry and the fundamental delay limit [19] will be discussed elsewhere.

Like the original point-to-point method, the basic idea of the current scheme is not restricted to optical fibers, but is applicable to any time-varying signal path of sufficient symmetry - in particular, it is well suited for free-space connections [18]. Similarly, dissemination of an optical carrier frequency is just a special case. The same principles apply for any periodic signal (such that we can define a phase $\Delta\phi_j$), including pulsed signals [23], radio frequency modulation (as recently implemented in [14]) and any signals modulated onto such carriers, or even the simultaneous dissemination of several frequencies [18]. In consequence, *synchronization* of many locations may be realized. Recently, endpoint synchronization was demonstrated using a slowly chirped optical frequency on a stabilized fiber link [27]. By combining this with the scheme above, we can also directly synchronize locations *along* the link, or in its vicinity.

A more general application follows without *stabilized* transmission paths, instead tapping a two-way time and frequency transfer (TWTFT) link [28]: we call this "eavesdropping" time (and frequency), see Fig. 5.

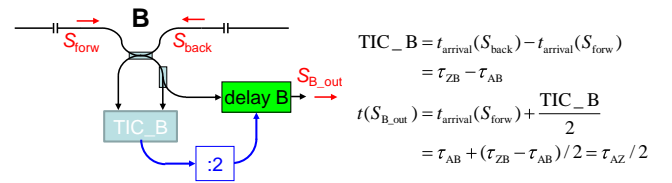

Fig. 5. Eavesdropping on a TWTFT link between points A and Z which simultaneously send signals $S_{forw}$ and $S_{back}$. At each extraction point, signal $S_{B\_out}$ exits the apparatus at time $\tau_{AZ}/2$.

Classic TWTFT, first demonstrated via the Telstar 1 satellite 50 years ago [29] and later implemented via fiber [28], uses counter-propagating signals sent and received by the two end-points A and Z. The time difference between sending and receiving is measured by time interval counters (TIC) at point A and point Z, to give the clock difference (simplified from [28,30]):

clock_A - clock_Z = (TIC_A - TIC_Z)/2 - ($\tau_{ZA} - \tau_{AZ}$)/2,

where $\tau_{AZ}$ is the signal delay between A and Z. For a symmetric link ($\tau_{ZA}=\tau_{AZ}$), this directly allows clock synchronization, i.e. clock_Z'=clock_A, and both endpoints can send a signal at the same time.

Now we tap these forward and backward travelling signals, at any point B (here chosen closer to A than Z), and measure the time difference for signal arrival TIC_B:= $t_{arrival}(S_{B\_back})$ - $t_{arrival}(S_{B\_forw})$ = $\tau_{ZB}$-$\tau_{AB}$. If we delay the forward extracted signal at B by ½TIC_B, it will exit our apparatus at point B at $t(S_{B\_out})$=($\tau_{AB}$+($\tau_{ZB}$-$\tau_{AB}$)/2)= $\tau_{AZ}$/2, in the timescale given by clock_A (and clock_Z'). Hence all points along such a link can be synchronized to each other, simply by eavesdropping, and share reference frequencies provided by A and Z. If they are additionally given the information of $\tau_{AZ}$, they may synchronize to the reference time scale of clock_A.

Implementation may use receiver modules of standard TWTFT-modems, as in point-to-point fiber based time-transfer [31,32], or newly developed electronic delay lines and time signal encoders/decoders [33].

In summary, we have presented new methods which – using very little extra instrumentation – make available at intermediate points along or near a transmission path the same timing or frequency signals that previously could only be delivered to its endpoints. Specifically, for applications requiring the highest precision, we demonstrated delivering an optical frequency to an intermediate point with relative instability $1\times10^{-17}$ (10 s) and $10^{-19}$ (10000 s). An improved branching design reaches users in the vicinity of a point-to-point fiber link. If a ring topology is used, the scheme enables monitored time/frequency dissemination to many users. The idea is applicable to any time-varying signal path, and we have outlined how to apply it e.g. to classic two-way time and frequency transfer links with intermediate access.

**Acknowledgement:** I am indebted to Fritz Riehle and Giorgio Santarelli for their indefatigable enthusiasm and openness to new ideas, to H. Schnatz for lab space and time to perform the experiments, and to the authors of [11] for lucid writing. I thank the Deutsche Forschungsgemeinschaft (SFB 407 and QUEST, Centre for Quantum Engineering and Space-Time Research), and the European Space Agency for financial support.